\documentclass[12pt,a4paper]{article}
\usepackage[latin1]{inputenc}
\usepackage[english]{babel}
\usepackage[T1]{fontenc}
\usepackage{ae,aecompl}
\usepackage{amssymb,amsmath,amsfonts}
\usepackage{ifpdf}
\ifpdf
  \usepackage[pdftex,a4paper,hmargin=28mm,vmargin=24mm]{geometry}
  \usepackage[pdftex,bookmarks=true]{hyperref}
\else
  \usepackage[dvips,a4paper,hmargin=28mm,vmargin=24mm]{geometry}
  \usepackage[dvips,bookmarks=true,pdfborder={0 0 112}]{hyperref}
\fi

\numberwithin{equation}{section}

\newcommand{\rd}{\mathrm{d}}
\newcommand{\dx}[1]{\rd x^{#1}}
\newcommand{\tr}{\mathrm{tr}}
\newcommand{\diag}{\mathrm{diag}}

\begin{document}
\begin{center}
{\Large Noncommutative gauge theory using covariant star product \\
\vspace{.3em}
defined between Lie-valued differential forms}\\
\vspace{1em}
M. Chaichian$^{a,b}$, M. Oksanen$^a$, A. Tureanu$^{a,b}$ and G. Zet$^c$\\
\vspace{1em}
\textit{${}^a$Department of Physics, University of Helsinki, P.O. Box 64,\\ FI-00014 
Helsinki, Finland\\
${}^b$Helsinki Institute of Physics, P.O. Box 64, FI-00014 Helsinki, Finland\\
${}^c$Department of Physics, The ``Gheorghe Asachi'' Technical University 
of Iasi,\\ Bd. D. Mangeron 67, 700050 Iasi, Romania}
\end{center}

\begin{abstract}
We develop an internal gauge theory using a covariant star product.
The space-time is a symplectic manifold endowed only with torsion
but no curvature. It is shown that, in order to assure the restrictions
imposed by the associativity property of the star product, the
torsion of the space-time has to be covariant constant.
An illustrative example is given and it is concluded that in
this case the conditions necessary to define a covariant star product
on a symplectic manifold completely determine its connection.
\end{abstract}

\section{Introduction}

Noncommutative gravity has been intensively studied in the
last years. One important motivation is the hope that such a theory
could offer the possibility to develop a quantum theory of gravity, or
at least to give an idea of how this could be achieved \cite{Szabo,Lizzi,Rivelles,Vassilevich:1,Aschieri:1,Aschieri:2}.
There are two major candidates to quantum gravity: string
theory \cite{Polchinski} and loop quantum gravity \cite{Rovelli}.
Noncommutative geometry and, in particular, gauge theory of gravity
are intimately connected with both these approaches and the overlaps
are considerable \cite{Lizzi}. String theory is one of the strongest
motivations for considering noncommutative space-time geometries
and noncommutative gravitation. It has been shown, for example,
that in the case when the end points of strings in a theory of open
strings are constrained to move on $D$ branes in a constant $B$-field
background and one considers the low-energy limit, then the full
dynamics of the theory is described by a gauge theory on a
noncommutative space-time \cite{Seiberg}. Recently, it has been
argued that the dynamics of the noncommutative gravity arising from
string theory \cite{Alvarez-Gaume} is much richer than some versions proposed for noncommutative gravity. It is suspected that the
reason for this is the noncovariance of the Moyal star product under
space-time diffeomorphisms. A geometrical approach to noncommutative
gravity, leading to a general theory of noncommutative Riemann
surfaces in which the problem of the frame dependence of the star
product is also recognized, has been proposed in \cite{Chaichian:1}
(for further developments, see \cite{Wang,Zhang}).

Now, one important problem is to develop a theory of gravity considering
curved noncommutative space-times. The main difficulty is that the
noncommutativity parameter $\theta^{\mu\nu}$ is
usually taken to be constant, which breaks the Lorentz invariance of
the commutation relations between coordinates [see \eqref{coordinate_cr} below],
and implicitly of any noncommutative field theory. One possible way to
solve this problem is to consider $\theta^{\mu\nu}$
depending on coordinates and use a covariant star product.
In \cite{McCurdy:1} such a product has been defined between
differential forms and the property of associativity was verified up
to the second order in $\theta^{\mu\nu}$.

In this paper we will adopt the covariant star product defined in
\cite{McCurdy:1} and extend the result to the case of Lie-algebra-valued
differential forms. We will follow the same procedure as in our
previous paper \cite{Chaichian:2}. But, in order to simplify the
expression of the covariant product and to give
an illustrative example, we will consider the case when the
noncommutative space-time is a symplectic manifold $M$ endowed
only with torsion (and no curvature). The restrictions imposed by such
a covariant star product requires also that the torsion is covariant
constant. The motivation for adopting such a manifold is that it allows
the construction of noncommutative teleparallel gravity (for the idea of
teleparallelism in gravity see \cite{Einstein}). It has been
shown that a very difficult problem like the definition of a tensorial
expression for the gravitational energy-momentum density can be solved
in teleparallel gravity \cite{Lucas,Aldrovandi}. This density is
conserved in covariant sense. It has also been argued that the 
quantization problem is much more convenient to handle in teleparallel 
gravity \cite{Andrade} than in general relativity, due to the possibility 
of decomposing torsion into irreducible pieces under the global Lorentz 
group.

Teleparallel gravity is also a very natural candidate for an
effective noncommutative field theory of gravitation \cite{Langmann}.
In addition it possesses many features which makes it particularly
well-suited for certain analyses. For instance it enables a pure
tensorial proof of the positivity of the energy in general relativity
\cite{Nester}, it yields a natural introduction of Ashtekar variables
\cite{Melke}, and makes it possible to study the torsion at quantum
level \cite{Okubo}, for example, in the gravitational coupling to
spinor fields.

On the other hand, we can try to apply the covariant star product to the
case when the space-time is a symplectic manifold which has only
curvature, but the torsion vanishes. Then, the restriction imposed by
the Jacobi identity for the Poisson bracket requires also the vanishing
curvature. The corresponding connection is flat symplectic and this
reduces drastically the applicability area of the covariant star product.
Of course, it is possible to have a manifold with both curvature
and torsion.

In Section~\ref{sec:star_product}, considering a manifold $M$ endowed
only with torsion, we give the definition of the covariant star
product between two arbitrary Lie-algebra-valued differential forms
and some of its properties. Then, the star bracket between such
differential forms is introduced and some examples are given.

Section~\ref{sec:NC_gauge_theory} is devoted to the noncommutative
internal gauge theory formulated with the new covariant star product.
The noncommutative Lie-algebra-valued gauge potential and the field
strength 2-form are defined and their gauge transformation laws are
established. It is shown that the field strength is gauge covariant
and satisfies a deformed Bianchi identity.

An illustrative example is presented in Section~\ref{sec:example}. It is shown that in
our simple example, the conditions necessary to define a covariant star
product on a symplectic manifold $M$ completely determine its
connection.

Section~\ref{sec:conclusions} is devoted to the discussion of the
results and to the interpretation of noncommutative gauge theory
formulated by using the covariant star product between Lie-algebra-valued differential forms on symplectic manifolds. Some other possible
applications of this covariant star product are also analyzed.

The Appendix~\ref{sec:appendix} contains a detailed verification of 
the associativity property of the covariant star product including only 
the torsion in its definition.

\section{Covariant star product}
\label{sec:star_product}

We consider a noncommutative space-time $M$ endowed with the
coordinates  $x^\mu,\, \mu = 0, 1, 2, 3$, satisfying the commutation relation
\begin{equation}
\left[ x^\mu, x^\nu \right]_\star = i \theta^{\mu\nu}(x) ,\label{coordinate_cr}
\end{equation}
where $\theta^{\mu\nu}(x) = - \theta^{\nu\mu}(x)$
is a Poisson bivector \cite{McCurdy:1}. The space-time is organized as a Poisson
manifold by introducing the Poisson bracket between two functions
$f(x)$ and $g(x)$ by
\begin{equation}
\{ f, g \} = \theta^{\mu\nu} \partial_\mu f \partial_\nu g .\label{Pb}
\end{equation}
In order that the Poisson bracket satisfies the Jacobi identity, the
bivector $\theta^{\mu\nu}(x)$ has to obey the condition \cite{Chu,McCurdy:2}
\begin{equation}
\theta^{\mu\sigma} \partial_\sigma \theta^{\nu\rho} + \theta^{\nu\sigma} \partial_\sigma \theta^{\rho\mu} + \theta^{\rho\sigma} \partial_\sigma \theta^{\mu\nu}= 0 .\label{Jacobi}
\end{equation}
If a Poisson bracket is defined on $M$, then $M$ is called
a Poisson manifold (see \cite{Chu} for mathematical details).

Suppose now that the bivector $\theta^{\mu\nu}(x)$
has an inverse $\omega_{\mu\nu}(x)$, i.e.
\begin{equation}
\theta^{\mu\rho} \omega_{\rho\nu} = \delta^\mu_\nu .\label{theta_invertible}
\end{equation}
If $\omega = \frac{1}{2} \omega_{\mu\nu} \dx{\mu} \wedge \dx{\nu}$
is nondegenerate $\left( \det \omega_{\mu\nu} \neq 0 \right)$ and closed  $\left( \rd \omega = 0 \right)$, then it is called a symplectic 2-form and $M$ is a symplectic
manifold. It can be verified that the condition $\rd \omega = 0$ is
equivalent with the equation \eqref{Jacobi} \cite{McCurdy:1,Chu,Tagliaferro}.
In this paper we will consider only the case when $M$ is symplectic.

Because the gauge theories involve Lie-valued differential forms such
as \linebreak
$A = A_\mu^{a}(x) T_{a} \dx{\mu} = A_\mu \dx{\mu},\, A_\mu =A_\mu^{a}(x) T_{a}$, where  $T_{a}$ are the infinitesimal generators of a symmetry group
$G$, we need to generalize the definition of the Poisson bracket
to differential forms and define then an associative star product for
such cases. These problems were solved in \cite{McCurdy:1,Chu,Tagliaferro}. In \cite{Chaichian:2}
we generalized these results to the case of Lie-algebra-valued
differential forms. This generalization has the effect that the
commutator of differential forms can be a commutator or an
anticommutator, depending on their degrees.

Assuming that $\theta^{\mu\nu}(x)$
is invertible, we can always write the Poisson bracket
$\left\{ x, dx \right\}$ in the form \cite{McCurdy:1,Chu,Tagliaferro}
\begin{equation}
\left\{ x^\mu, \dx{\nu} \right\} = - \theta^{\mu\rho} \Gamma^\nu_{\rho\sigma} \dx{\sigma} ,
\end{equation}
where $\Gamma^\nu_{\rho\sigma}$ are some functions of $x$ transforming
like a connection under general coordinate transformations. As
$\Gamma^\nu_{\rho\sigma}$ is generally not symmetric, one can use
the connection 1-forms
\begin{equation}
\tilde{\Gamma}^\mu_\nu = \Gamma^\mu_{\nu\rho} \dx{\rho}
,\quad
\Gamma^\mu_\nu = \dx{\rho} \Gamma^\mu_{\rho\nu}
\end{equation}
to define two kinds of covariant derivatives $\tilde{\nabla}$ and $\nabla$, respectively.
The curvatures for these two connections are
\begin{align}
\tilde{R}^\nu_{\lambda\rho\sigma} &= \partial_\rho \Gamma^\nu_{\lambda\sigma} - \partial_\sigma \Gamma^\nu_{\lambda\rho} + \Gamma^\nu_{\tau\rho} \Gamma^\tau_{\lambda\sigma} - \Gamma^\nu_{\tau\sigma} \Gamma^\tau_{\lambda\rho} ,\label{tildeR}\\
R^\nu_{\lambda\rho\sigma} &= \partial_\rho \Gamma^\nu_{\sigma\lambda} - \partial_\sigma \Gamma^\nu_{\rho\lambda} + \Gamma^\nu_{\rho\tau} \Gamma^\tau_{\sigma\lambda} - \Gamma^\nu_{\sigma\tau} \Gamma^\tau_{\rho\lambda} .\label{R}
\end{align}
Because the connection coefficients $\Gamma^\rho_{\mu\nu}$
are not symmetric
$\left( \Gamma^\rho_{\mu\nu} \neq \Gamma^\rho_{\nu\mu} \right)$
the symplectic manifold $M$ has also a torsion defined as usual \cite{McCurdy:1}
\begin{equation}
T^\rho_{\mu\nu} = \Gamma^\rho_{\mu\nu} - \Gamma^\rho_{\nu\mu} .
\end{equation}

The connection  ${\nabla }$ satisfies the identity \cite{McCurdy:1}
\begin{equation}
\left[ \nabla_\mu, \nabla_\nu \right] \alpha = - R^\sigma_{\rho\mu\nu} \dx{\rho} \wedge i_\sigma \alpha - T^\rho_{\mu\nu} \nabla_\rho \alpha ,\label{cocd}
\end{equation}
and an analogous formula applies for $\tilde{\nabla}$. Here,
$\alpha$ is an arbitrary differential $k$-form
\begin{equation}
\alpha = \frac{1}{k!} \alpha_{\mu_1 \cdots \mu_k} \dx{\mu_1} \wedge \cdots \wedge \dx{\mu_k}
\end{equation}
and  $i_\sigma \alpha$ denotes the interior product which maps the
$k$-form $\alpha$ into a $(k-1)$-form
\begin{equation}
i_\sigma \alpha = \frac{1}{(k-1)!} \alpha_{\sigma \mu_2 \cdots \mu_k} \dx{\mu_2} \wedge \cdots \wedge \dx{\mu_k} .
\end{equation}

It has been proven that in order for the Poisson bracket to satisfy the
Leibniz rule
\begin{equation}
\rd \{ f, g \} = \{ \rd f, g \} + \{ f, \rd g \}
\end{equation}
the bivector $\theta^{\mu\nu}(x)$ has to obey the property \cite{McCurdy:1}
\begin{equation}
\tilde{\nabla}_\rho \theta^{\mu\nu} = \partial_\rho \theta^{\mu\nu} + \Gamma^\mu_{\sigma\rho} \theta^{\sigma\nu} + \Gamma^\nu_{\sigma\rho} \theta^{\mu\sigma} = 0 .\label{tildenabla_symplectic}
\end{equation}
Thus  $\theta^{\mu\nu}$ is covariant constant under $\tilde{\nabla }$, and $\tilde{\nabla}$ is an almost symplectic connection. One can use the
Leibniz condition \eqref{tildenabla_symplectic} together with the Jacobi identity for the
Poisson bivector $\theta^{\mu\nu}$ to obtain the cyclic relation for torsion
\begin{equation}
\sum_{(\mu, \nu, \rho)} \theta^{\mu\sigma} \theta^{\nu\lambda} T^\rho_{\sigma\lambda} = 0 .\label{T_cyclic}
\end{equation}
Note that while this relation shows that a torsion-free connection
identically satisfies the property \eqref{T_cyclic}, the Jacobi identity does not
require the connection to be torsionless. Also note that \eqref{tildenabla_symplectic}
and the Jacobi identity for the Poisson bivector can be combined to obtain the
following cyclic relation:
\begin{equation}
\sum_{(\mu, \nu, \rho)} \theta^{\mu\sigma} \nabla_\sigma \theta^{\nu\rho} = 0 .\label{Jacobi_covariant}
\end{equation}
If in addition to  $\tilde{\nabla}_\rho \theta^{\mu\nu} = 0$, one imposes
$\nabla_\rho \theta^{\mu\nu} = 0$, the torsion vanishes,
$T^\rho_{\mu\nu} = 0$, and there is only one covariant derivative
$\nabla = \tilde{\nabla}$. In this paper, we do not require that
$\nabla_\rho \theta^{\mu\nu} = 0$.

Using the graded product rule, one arrives at the following general
expression of the Poisson bracket between differential forms \cite{McCurdy:1,Tagliaferro}
\begin{equation}
\{ \alpha, \beta \} = \theta^{\mu\nu} \nabla_\mu \alpha \wedge \nabla_\nu \beta + (-1)^{|\alpha|} \tilde{R}^{\mu\nu} \wedge (i_\mu \alpha) \wedge (i_\nu \beta) ,\label{Pb_form}
\end{equation}
where $|\alpha|$ is the degree of the differential form $\alpha$, and
\begin{equation}
\tilde{R}^{\mu\nu} = \frac{1}{2} \tilde{R}^{\mu\nu}_{\rho\sigma} \dx{\rho} \wedge \dx{\sigma}
, \quad
\tilde{R}^{\mu\nu}_{\rho\sigma} = \theta^{\mu\lambda} \tilde{R}^\nu_{\lambda\rho\sigma}
\end{equation}
In order that \eqref{Pb_form} satisfies the graded Jacobi identity
\begin{equation}
\{ \alpha, \{ \beta, \gamma \} \} + (-1)^{|\alpha| (|\beta|+|\gamma|)} \{ \beta, \{ \gamma, \alpha \} \} + (-1)^{|\gamma| (|\alpha|+|\beta|)} \{ \gamma, \{ \alpha, \beta \} \} = 0 ,\label{Jacobi_Pb}
\end{equation}
the connection $\Gamma^\rho_{\mu\nu}$ must obey the following
additional conditions \cite{McCurdy:1}
\begin{gather}
R^\nu_{\lambda\rho\sigma} = 0 ,\label{R_zero}\\
\nabla_\lambda \tilde{R}^{\mu\nu}_{\rho\sigma} = 0 .\label{tildeR_covariant_constant}
\end{gather}

A covariant star product between arbitrary differential forms has been
defined recently in \cite{McCurdy:1} having the general form
\begin{equation}
\alpha \star \beta = \alpha \wedge \beta + \sum_{n=1}^\infty \left(\frac{i\hbar}{2}\right)^n C_n \left( \alpha, \beta \right), \label{star_product}
\end{equation}
where $C_n (\alpha, \beta)$ are bilinear differential operators satisfying
the generalized Moyal symmetry \cite{Tagliaferro}
\begin{equation}
C_n (\alpha, \beta) = (-1)^{|\alpha| |\beta| + n} C_n (\beta, \alpha). \label{Moyal_symmetry}
\end{equation}
The operator $C_1$ coincides with the Poisson bracket, i.e.
$C_1 (\alpha, \beta) = \{ \alpha, \beta \}$.
An expression for $C_2 (\alpha, \beta)$ has been obtained
also in \cite{McCurdy:1} so that the star product \eqref{star_product} satisfies the
property of associativity
\begin{equation}
(\alpha \star \beta) \star \gamma = \alpha \star (\beta \star \gamma) .\label{associativity}
\end{equation}

In this paper we consider the case when the symplectic manifold $M$ has
only torsion, i.e. in addition to the necessary constraints \eqref{tildenabla_symplectic}, \eqref{R_zero} and \eqref{tildeR_covariant_constant} we require
\begin{equation}
\tilde{R}^\sigma_{\mu\nu\rho} = 0 .\label{tildeR_zero}
\end{equation}
Since the curvature $R^\sigma_{\rho\mu\nu}$ vanishes \eqref{R_zero},
one obtains the following relation between
the curvature $\tilde{R}$ and the torsion $T$ \cite{McCurdy:1}
\begin{equation}
\tilde{R}^\sigma_{\mu\nu\rho} = \nabla_\mu T^\sigma_{\nu\rho} .
\end{equation}
This relation shows that the condition \eqref{tildeR_zero} requires
that the torsion $T^\sigma_{\nu\rho}$ is covariant constant, i.e.
\begin{equation}
\nabla_\mu T^\sigma_{\nu\rho} = 0 .\label{nabla_T_zero}
\end{equation}
Therefore, if the torsion is covariant constant, the symplectic manifold
$M$ has only torsion but not curvature.

For such a symplectic manifold, the bilinear differential operators
$C_1 (\alpha, \beta)$ and $C_2 (\alpha, \beta)$ in the star
product \eqref{star_product} proposed in \cite{McCurdy:1} reduce to the simpler forms
\begin{align}
C_1 (\alpha, \beta) &= \{ \alpha, \beta \} = \theta^{\mu\nu} \nabla_\mu \alpha \wedge \nabla_\nu \beta ,\label{C_1}\\
C_2 (\alpha, \beta) &= \frac{1}{2} \theta^{\mu\nu} \theta^{\rho\sigma} \nabla_\mu \nabla_\rho \alpha \wedge \nabla_\nu \nabla_\sigma \beta
+ \frac{1}{3} \left( \theta^{\mu\sigma} \nabla_\sigma \theta^{\nu\rho} + \frac{1}{2} \theta^{\nu\sigma} \theta^{\rho\lambda} T^\mu_{\sigma\lambda} \right) \label{C_2}\\
&\quad\times \left( \nabla_\mu \nabla_\nu \alpha \wedge \nabla_\rho \beta - \nabla_\nu \alpha \wedge \nabla_\mu \nabla_\rho \beta \right) .\nonumber
\end{align}
We can verify that the covariant star product with torsion defined in
\eqref{C_1}--\eqref{C_2} is associative [see the Appendix~\ref{sec:appendix}].
In the next section we apply this covariant star product in order to develop a noncommutative internal gauge theory.

\section{Noncommutative gauge theory}
\label{sec:NC_gauge_theory}

Let us consider the internal symmetry group $G$ and develop a
noncommutative gauge theory on the symplectic manifold $M$ endowed with the covariant star product (with torsion) defined above.
We proceed as in \cite{Chaichian:2}, but considering the star product defined in
\eqref{C_1}--\eqref{C_2}. This product differs from that used in
\cite{Chaichian:2}:
\begin{itemize}
\item The curvature  $\tilde{R}^\sigma_{\mu\nu\rho}$
is supposed here to vanish as well as $R^\sigma_{\mu\nu\rho}$;
\item The ordinary derivative  $\partial_\sigma \theta^{\nu\rho}$ (see
(2.17) in \cite{Chaichian:2}) is replaced with the covariant derivative
$\nabla_\sigma \theta^{\nu\rho}$;
\item In \eqref{C_2} it appears an additional term
$\frac{1}{2} \theta^{\nu\sigma} \theta^{\rho\lambda} T^\mu_{\sigma\lambda}$
compared with previous version (see \cite{McCurdy:1} for details, and
also \cite{Tagliaferro,McCurdy:2} for other aspects).
\end{itemize}
The results given in \cite{Chaichian:2} apply with the corresponding changes
mentioned above. Before presenting them we make some observations on other
possible applications of the covariant star product.
\begin{enumerate}
\item It will be interesting to see if the Seiberg-Witten map can be
generalized to the case when the ordinary derivatives are replaced with
the covariant derivatives and the Moyal star product is replaced by the covariant
one.
\item We can consider that the symplectic manifold $M$ is associated
to a gauge theory of gravitation with Poincar\'e group  $P$ as local
symmetry (see \cite{Blagojevic,Chaichian:3,Chaichian:4} for
notations and definitions) and using the covariant star product as in
\cite{Chaichian:2}. In this case we introduce the Poincar\'e
gauge fields  $e^a_{\phantom{a}\mu}$ (tetrads) and $\omega_\mu^{ab}$
(spin connection) and then we define the covariant derivative as
\begin{equation}
\nabla_\mu = \partial_\mu - \frac{1}{2} \omega_\mu^{ab} \Sigma_{ab} .
\end{equation}
It can be shown that by imposing the tetrad postulate \cite{Blagojevic}
\begin{equation}
\nabla_\mu e^a_{\phantom{a}\nu} - \Gamma^\rho_{\mu\nu} e^a_{\phantom{a}\rho} = 0
\end{equation}
one introduces the connection $\Gamma^\rho_{\mu\nu}$
in the Poincar\'e gauge theory and the strength tensors
$F^{ab}_{\phantom{ab}\mu\nu},\, F^a_{\phantom{a}\mu\nu}$
determine the curvature and torsion of $M$
\begin{equation}
R^{\rho\sigma}_{\mu\nu} = F^{ab}_{\phantom{ab}\mu\nu} \bar{e}_a^{\phantom{a}\rho} \bar{e}_b^{\phantom{b}\sigma}
,\quad
T^\rho_{\mu\nu} = F^a_{\phantom{a}\mu\nu} \bar{e}_a^{\phantom{a}\rho} ,
\end{equation}
where  $\bar{e}_a^{\phantom{a}\rho}$ denote the inverse of
$e^a_{\phantom{a}\mu}$, i.e.
\begin{equation}
\bar{e}_a^{\phantom{a}\rho} e^a_{\phantom{a}\sigma} = \delta^\rho_\sigma
,\quad
\bar{e}_a^{\phantom{a}\rho} e^b_{\phantom{b}\rho} = \delta_a^b .
\end{equation}

Now, let us suppose that we develop an internal gauge theory with the
symmetry group $G$ on the symplectic manifold $M$. It is
very important to remark that making the minimal prescription
$\partial_\mu \rightarrow \nabla_\mu$ the strength tensor $F_{\mu\nu}$
of the internal gauge fields $A_\mu = A_\mu^a T_a$ must be
written as
\begin{equation}
F_{\mu\nu} = \nabla_\mu A_\nu - \nabla_\nu A_\mu - i \left[ A_\mu, A_\nu \right] + A_\rho T^\rho_{\mu\nu}, \label{F}
\end{equation}
in order to assure its covariance both under Poincar\'e group $P$ and
internal group $G$. The gauge invariance under the gauge
transformations of $G$ becomes quite clear if we observe that
the expression \eqref{F} is identical with the usual one. Indeed, using the
definitions of the covariant derivative  $\nabla_\mu$
and torsion $T^\rho_{\mu\nu}$, we obtain
\begin{equation}
\begin{split}
F_{\mu\nu} &= \partial_\mu A_\nu - A_\rho \Gamma^\rho_{\mu\nu} - \partial_\nu A_\mu + A_\rho \Gamma^\rho_{\nu\mu} - i \left[ A_\mu, A_\nu \right] + A_\rho \left( \Gamma^\rho_{\mu\nu} - \Gamma^\rho_{\nu\mu} \right) \\
&= \partial_\mu A_\nu - \partial_\nu A_\mu - i \left[ A_\mu, A_\nu \right] .
\end{split}
\end{equation}

Also, because the components of the gauge parameter $\lambda = \lambda^a T_a$
are considered as functions, we can write the gauge transformations as
\begin{equation}
\delta A_\mu = \nabla_\mu \lambda - i \left[ A_\mu, \lambda \right]
,\quad
\nabla_\mu \lambda = \partial_\mu \lambda .\label{delta_A}
\end{equation}
In what follows, we will use these expressions \eqref{F} and \eqref{delta_A} in order
to show their invariances explicitly.
\end{enumerate}

Suppose now that we have an internal gauge group $G$ whose
infinitesimal generators  $T_a$ satisfy the algebra
\begin{equation}
\left[ T_a, T_b \right] = i f_{ab}^c T_c ,\quad a, b, c = 1, 2, \ldots, m ,\label{T_Lie_algebra}
\end{equation}
with the structure constants $f_{ab}^c = - f_{ba}^c$
and that the Lie-algebra-valued infinitesimal parameter is
\begin{equation}
\hat{\lambda} = \hat{\lambda}^a T_a .
\end{equation}
We use the hat symbol ``$\hat{\phantom{a}}$'' to denote the
noncommutative quantities of our gauge theory. The parameter
$\hat{\lambda}$ is a zero-form, i.e.  $\hat{\lambda}^a$ are functions
of the coordinates $x^\mu$ on the symplectic manifold $M$.

Now, we define the  gauge transformation of the noncommutative 
Lie-valued gauge potential
\begin{equation}
\hat{A} = \hat{A}_\mu^a (x) T_a \dx{\mu} = \hat{A}_\mu \dx{\mu}
,\quad
\hat{A}_\mu = \hat{A}_\mu^a (x) T_a ,
\end{equation}
by
\begin{equation}
\hat{\delta} \hat{A} = \rd \hat{\lambda} - i \left[ \hat{A}, \hat{\lambda} \right]_\star .\label{delta_A_nc}
\end{equation}
Here we consider the following formula for the commutator $[\alpha, \beta]_\star$
of two arbitrary differential forms  $\alpha$ and $\beta$
\begin{equation}
[\alpha, \beta]_\star = \alpha \star \beta - (-1)^{|\alpha| |\beta|} \beta \star \alpha .\label{star_commutator}
\end{equation}
Then, using the definition \eqref{star_product} of the star product,
we can write \eqref{delta_A_nc} as
\begin{equation}
\hat{\delta} \hat{A}^a = \rd \hat{\lambda}^a + f^a_{bc} \hat{A}^b \hat{\lambda}^c + \frac{\hbar}{2} d^a_{bc} C_1 \left( \hat{A}^b, \hat{\lambda}^c \right) - \frac{\hbar^2}{4} f^a_{bc} C_2 \left( \hat{A}^b, \hat{\lambda}^c \right) + O\left( \hbar^3 \right) ,
\end{equation}
where we noted $\left\{ T_a, T_b \right\} = d_{ab}^c T_c$.
In fact, this notation is valid if the Lie algebra closes also for the 
anticommutator, as it happens, for example, in the case of unitary
groups. In general, the commutators like
$\left[ \hat{A}, \hat{\lambda} \right]_\star$ take
values in the enveloping algebra \cite{Garcia-Compean}. Therefore, the
gauge field $\hat{A}$ and the parameter $\hat{\lambda}$ take
values in this algebra. Let us write for instance
$\hat{A} = \hat{A}^I T_I$ and $\hat{\lambda} = \hat{\lambda}^I T_I$,
then
\begin{equation*}
\left[ \hat{A}, \hat{\lambda} \right]_\star = \frac{1}{2} \left\{ \hat{A}^I, \hat{\lambda}^J \right\}_\star [T_I, T_J] + \frac{1}{2} \left[ \hat{A}^I, \hat{\lambda}^J \right]_\star \{T_I, T_J\} .
\end{equation*}
Thus, all products of the generators  $T_I$ will be necessary in
order to close the enveloping algebra. Its structure can be obtained by
successively computing the commutators and anticommutators starting
from the generators of Lie algebra, until it closes \cite{Garcia-Compean},
\begin{equation*}
\left[ T_I, T_J \right] = i f_{IJ}^K T_K
,\quad
\left\{ T_I, T_J \right\} = d_{IJ}^K T_K .
\end{equation*}
Therefore, in our above notations and in what follows we understand this
structure in general.

The operators $C_n$ of the star product are defined similarly for
noncommutative differential forms like $\hat{A}^a$ as for commutative ones.
In particular $C_1 \left( \hat{A}^b, \hat{\lambda}^c \right)$ and
$C_2 \left( \hat{A}^b, \hat{\lambda}^c \right)$
are given by \eqref{C_1}--\eqref{C_2}.
Here the covariant derivative concerns the space-time manifold $M$, not the gauge group $G$, so we use the definition
\begin{equation}
\nabla_\mu \hat{A}^a = \left( \partial_\mu \hat{A}^a_\nu - \Gamma^\rho_{\mu\nu} \hat{A}^a_\rho \right) \dx{\nu} \equiv \left( \nabla_\mu \hat{A}^a_\nu \right) \dx{\nu} .
\end{equation}

We define also the curvature 2-form $\hat{F}$ of the gauge
potentials by
\begin{equation}
\hat{F} = \frac{1}{2} \dx{\mu} \wedge \dx{\nu} \hat{F}_{\mu\nu} = \rd \hat{A} - \frac{i}{2} \left[ \hat{A}, \hat{A} \right] .\label{F_nc}
\end{equation}
Then, using the definition \eqref{star_product} of the star product and the property
\eqref{Moyal_symmetry} of the operators
$C_n \left( \alpha^a, \beta^b\right)$, we obtain from \eqref{F_nc}
\begin{equation}
\hat{F}^a = \rd \hat{A}^a + \frac{1}{2} f^a_{bc} \hat{A}^b \wedge \hat{A}^c + \frac{1}{2} \frac{\hbar}{2} d^a_{bc} C_1 \left( \hat{A}^b, \hat{A}^c \right) - \frac{1}{2} \frac{\hbar^2}{4} f^a_{bc} C_2 \left( \hat{A}^b, \hat{A}^c \right) + O\left( \hbar^3 \right)
\end{equation}
More explicitly, in terms of components we have
\begin{equation}
\hat{F}^a_{\mu\nu} = \nabla_\mu \hat{A}^a_\nu - \nabla_\nu \hat{A}^a_\mu  + f^a_{bc} \hat{A}^b_\mu \hat{A}^c_\nu
+ \hat{A}^a_\rho T^\rho_{\mu\nu} + \frac{\hbar}{2} d^a_{bc} C_1 \left( \hat{A}^b_\mu, \hat{A}^c_\nu \right) - \frac{\hbar^2}{4} f^a_{bc} C_2 \left( \hat{A}^b_\mu, \hat{A}^c_\nu \right) + O\left( \hbar^3 \right) ,
\end{equation}
where we used the definition
$C_n \left( \hat{A}^b, \hat{A}^c \right) = C_n \left( \hat{A}^b_\mu, \hat{A}^c_\nu \right) \dx{\mu} \wedge \dx{\nu}$ with
\begin{align}
C_1 \left( \hat{A}^b_\mu, \hat{A}^c_\nu \right) &= \left\{ \hat{A}^b_\mu, \hat{A}^c_\nu \right\} = \theta^{\rho\sigma} \nabla_\rho \hat{A}^b_\mu \nabla_\sigma \hat{A}^c_\nu ,\\
C_2 \left( \hat{A}^b_\mu, \hat{A}^c_\nu \right) &= \frac{1}{2} \theta^{\rho\sigma} \theta^{\lambda\tau} \nabla_\rho \nabla_\lambda \hat{A}^b_\mu \nabla_\sigma \nabla_\tau \hat{A}^c_\nu
+ \frac{1}{3} \left( \theta^{\rho\tau} \nabla_\tau \theta^{\sigma\lambda} + \frac{1}{2} \theta^{\sigma\tau} \theta^{\lambda\phi} T^\rho_{\tau\phi} \right) \\
&\quad\times \left( \nabla_\rho \nabla_\sigma \hat{A}^b_\mu \nabla_\lambda \hat{A}^c_\nu - \nabla_\sigma \hat{A}^b_\mu \nabla_\rho \nabla_\lambda \hat{A}^c_\nu \right) .\nonumber
\end{align}

Under the gauge transformation \eqref{delta_A_nc} the curvature 2-form
$\hat{F}$ transforms as
\begin{equation}
\hat{\delta} \hat{F} = i \left[ \hat{\lambda}, \hat{F} \right]_\star ,\label{delta_F_nc}
\end{equation}
where we used the Leibniz rule
\begin{equation}
\rd \left( \hat{\alpha} \star \hat{\beta} \right) = \rd \hat{\alpha} \star \hat{\beta} + (-1)^{|\alpha|} \hat{\alpha} \star \rd \hat{\beta}
\end{equation}
which we admit to be valid to all orders in $\hbar$.
The Leibniz rule was verified only in the first order.
To second order the proof is very cumbersome. We believe, however,
motivated by the associativity of the proposed star product, that
the Leibniz rule is valid to all orders. This issue remains a
challenge to be proven.
In terms of the components
\eqref{delta_F_nc} gives
\begin{equation}
\hat{\delta} \hat{F}^a = f^a_{bc} \hat{F}^b \hat{\lambda}^c +
\frac{\hbar}{2} d^a_{bc} C_1 \left( \hat{F}^b, \hat{\lambda}^c
\right) - \frac{\hbar^2}{4} f^a_{bc} C_2 \left( \hat{F}^b,
\hat{\lambda}^c \right) + O\left( \hbar^3 \right)
.\label{delta_F_nc_comp}
\end{equation}
In the zeroth order, the formula \eqref{delta_F_nc_comp} reproduces therefore the result
of the commutative gauge theory
\begin{equation}
\delta F^a_{\mu\nu} = f^a_{bc} F^b_{\mu\nu} \lambda^c \ \Leftrightarrow\ \delta F = i [\lambda, F] .
\end{equation}

Using again the Leibniz rule, we obtain the deformed Bianchi identity
\begin{equation}
\rd \hat{F} - i \left[ \hat{A}, \hat{F} \right]_\star = 0 .\label{Bianchi_nc}
\end{equation}
If we apply the definition \eqref{star_commutator} of the star commutator, we obtain
\begin{equation}
\rd \hat{F} + i \left[ \hat{F}, \hat{A} \right] = \left[ \frac{\hbar}{2} d^a_{bc} C_1 \left( \hat{F}^b, \hat{A}^c \right) - \frac{\hbar^2}{4} f^a_{bc} C_2 \left( \hat{F}^b, \hat{A}^c \right) \right] T_a + O\left( \hbar^3 \right) ,\label{Bianchi_nc_expanded}
\end{equation}
or in terms of components
\begin{equation}
\rd \hat{F}^a - f^a_{bc} \hat{F}^b \wedge \hat{A}^c = \frac{\hbar}{2} d^a_{bc} C_1 \left( \hat{F}^b, \hat{A}^c \right) - \frac{\hbar^2}{4} f^a_{bc} C_2 \left( \hat{F}^b, \hat{A}^c \right) + O\left( \hbar^3 \right) .
\end{equation}
We remark that in zeroth order we obtain from \eqref{Bianchi_nc_expanded}
the usual Bianchi identity
\begin{equation}
\rd F + i [F, A] = 0 .
\end{equation}
In addition, if the gauge group is  ${U\left(1\right)}$, the Bianchi
identity \eqref{Bianchi_nc} becomes
\begin{equation}
\rd \hat{F} = \hbar C_1 \left( \hat{A}, \hat{F} \right) + O\left( \hbar^3 \right) .
\end{equation}
This result is also in accord with that of \cite{Chu}.

Having established the previous results, we can construct a noncommutative Yang-Mills (NCYM) action. We will consider the case when the gauge group is $U(N)$. Let $G^{\mu\nu}$ be a metric on the noncommutative space-time $M$ \cite{Chaichian:2}. We suppose that the metric $G^{\mu\nu}$ belongs to the adjoint representation of $U(1) \subset U(N)$ in the sense that $G^{\mu\nu} = G^{\mu\nu} I$, where $I$ is the unity matrix of $U(N)$ in this representation. Therefore, we consider the components of $G^{\mu\nu}$ as Lie-algebra-valued zero-forms. The covariant derivative of the metric $G^{\mu\nu}$ is
\begin{equation}
\nabla_\mu G^{\nu\rho} = \partial_\mu G^{\nu\rho} + G^{\nu\sigma}\Gamma^\rho_{\mu\sigma} + \Gamma^\nu_{\mu\sigma}G^{\sigma\rho} .\label{nabla_G}
\end{equation}

If $G^{\mu\nu}$ is not constant, we have to modify it to be a covariant metric $\hat{G}^{\mu\nu}$ for the (NCYM) action \cite{Ho:1}, so that it transforms like $\hat{F}$ [see \eqref{delta_F_nc}]
\begin{equation}
\hat{\delta} \hat{G}^{\mu\nu} = i \left[ \hat{\lambda}, \hat{G}^{\mu\nu} \right]_\star ,\label{delta_G_nc}
\end{equation}
Then, using the definition \eqref{star_commutator} for the star commutator, we obtain from \eqref{delta_G_nc}
\begin{equation}
\hat{\delta}\hat{G}^{\mu\nu} = \theta^{\rho\sigma}\nabla_\rho \hat{G}^{\mu\nu} \partial_\sigma \hat{\lambda} + O\left( \hbar^3 \right) .
\end{equation}
We can use the Seiberg-Witten map with the covariant star product for a field, which is in the adjoint representation (as we consider $G^{\mu\nu}$ to be), in order to obtain \cite{Chaichian:5}
\begin{equation}
\hat{G}^{\mu\nu} = G^{\mu\nu} - A^0_\rho \theta^{\rho\sigma} \nabla_\sigma G^{\mu\nu} + O\left( \hbar^3 \right) ,
\end{equation}
where $A^0_\mu$ is the gauge field in the $U(1)$ sector of $U(N)$.

In order to construct the NCYM action for the gauge fields $A_\mu^a(x), \mu=1,2,3,0, a=0,1,2,\ldots,N^2-1$, we use the definition for the integration $\langle f\rangle$ of a function $f$ (or of any other quantity) over the noncommutative space $M$ as (see \cite{Ho:1} for details)
\begin{equation}
\langle\cdot\rangle = \int\dx{4} \left|Pf(B)\right| (\cdot) ,\label{integral}
\end{equation}
where $B=\theta^{-1}$ and $Pf(B)$ denotes the Pfaffian of $B$, i.e. $\left|Pf(B)\right|=\sqrt{\det(B)}$. The notation has a connection with the very important result that for a $D$-brane in a $B$ field background (with $B$ constant or nonconstant), its low-energy effective theory lives on a noncommutative space-time with the Poisson structure \cite{Ho:1,Ho:2,Cattaneo}. More exactly, it has been shown that the metric $G$ introduced on the Poisson manifold $M$ is related to the metric $g$ appearing in the fundamental string (open or closed) action by $G=-B^{-1} g B^{-1}$ \cite{Seiberg,Ho:1,Ho:2}.

Now, we define the NCYM action by (see \cite{Chaichian:2,Ho:1})
\begin{equation}
\hat{S}_\mathrm{NCYM} = -\frac{1}{2g_c^2} \left\langle \tr \left( \hat{G}\star\hat{F}\star\hat{G}\star\hat{F} \right) \right\rangle = -\frac{1}{4g_c^2} \left\langle \tr \left( \hat{G}^{\mu\nu}\star\hat{F}_{\nu\rho}\star\hat{G}^{\rho\sigma}\star\hat{F}_{\sigma\mu} \right) \right\rangle ,\label{S_NCYM}
\end{equation}
where $g_c$ is the Yang-Mills gauge coupling constant, and we have used the normalization property
\begin{equation}
\tr\left( T_a T_b \right) = \frac{1}{2} \delta_{ab} I .
\end{equation}

Using the properties of gauge covariance \eqref{delta_F_nc} and \eqref{delta_G_nc} for $\hat{F}$ and $\hat{G}$, respectively, we obtain
\begin{equation}
\hat{\delta}\hat{S}_\mathrm{NCYM} = -\frac{\hbar}{4g_c^2} \left\langle C_1 \left( \tr \left( \hat{G}\hat{F}\hat{G}\hat{F} \right), \hat{\lambda} \right) \right\rangle + O\left( \hbar^3 \right) .
\end{equation}
Since the integral \eqref{integral} is cyclic in the Poisson limit \cite{Ho:1}, the integral of the Poisson bracket vanishes, i.e. $\langle C_1(f, h) \rangle=0$ for any integrable functions $f$ and $h$, and thus (3.36) becomes
\begin{equation}
\hat{\delta}\hat{S}_\mathrm{NCYM} = 0 + O\left( \hbar^3 \right) .
\end{equation}
Therefore, the action $\hat{S}_\mathrm{NCYM}$ is invariant up to the
second order in $\hbar$. The expression \eqref{S_NCYM} of the action
can be further simplified as \cite{Chaichian:2,Ho:1}
\begin{equation}
\hat{S}_\mathrm{NCYM} = -\frac{1}{2g_c^2} \left\langle \tr \left( \hat{G}\hat{F}\hat{G}\hat{F} \right) \right\rangle + O\left( \hbar^3 \right) = -\frac{1}{4g_c^2} \left\langle \tr \left( \hat{G}^{\mu\nu}\hat{F}_{\nu\rho}\hat{G}^{\rho\sigma}\hat{F}_{\sigma\mu} \right) \right\rangle + O\left( \hbar^3 \right) ,\label{S_NCYM_simp}
\end{equation}

Using the previous results we can obtain solutions for the
noncommutative gauge field equations. An example is given in Section
\ref{sec:example}, using the symplectic manifold $M$ endowed with a
covariantly constant torsion.

We can add fields into our noncommutative gauge model in the usual way. As an example, we mention the case when the noncommutative $U(N)$ gauge theory is coupled to a Higgs multiplet $\hat{\Phi}(x)=\hat{\Phi}^a T_a$ in the adjoint representation. The action integral for $\hat{\Phi}(x)$ is \cite{Cieri}
\begin{equation}
\hat{S}_\mathrm{HIGGS} = -\frac{1}{4g^2} \left\langle \tr \left( \hat{D}_\mu \hat{\Phi}\star\hat{G}^{\mu\nu}\star\hat{D}_\nu \hat{\Phi} \right) \right\rangle ,
\end{equation}
where
\begin{equation}
\hat{D}_\mu \hat{\Phi} = \partial_\mu \hat{\Phi} - ig\left[ \hat{\Phi}, \hat{A}_\mu \right]_\star
\end{equation}
is the noncommutative gauge covariant derivative. Because this derivative is gauge covariant, in the sense
\begin{equation}
\hat{\delta}\left(\hat{D}_\mu \hat{\Phi}\right) = i\left[ \hat{\lambda}, \hat{D}_\mu \hat{\Phi} \right]_\star ,
\end{equation}
the action $\hat{S}_\mathrm{HIGGS}$ is invariant as well as
$\hat{S}_\mathrm{NCYM}$ up to the second order in $\hbar$. The
action of the noncommutative $U(N)$ coupled to the Higgs multiplet
$\hat{\Phi}(x)$ reads
\begin{equation}
\hat{S}_\mathrm{NC} = -\frac{1}{4g^2} \left\langle \tr \left( \hat{G}^{\mu\nu}\star\hat{F}_{\nu\rho}\star\hat{G}^{\rho\sigma}\star\hat{F}_{\sigma\mu} + \hat{D}_\mu \hat{\Phi}\star\hat{G}^{\mu\nu}\star\hat{D}_\nu \hat{\Phi} \right) \right\rangle .\label{S_NC}
\end{equation}
This action can be used for obtaining solutions of the noncommutative version of the Yang-Mills Higgs model by using the covariant star product on the symplectic manifold $M$, as an extension of the results of \cite{Cieri}, where the usual Moyal star product is used.

\section{An illustrative example}
\label{sec:example}

As a very simple example we consider the Poincar\'e gauge theory to
construct the manifold $M$. Then, suppose that we have the
gauge fields $e^a_{\phantom{a}\mu}$ and fix the gauge
$\omega_\mu^{ab} = 0$ \cite{Zet}. We define the connection coefficients
\begin{equation}
\Gamma^\rho_{\mu\nu} = \bar{e}_a^{\phantom{a}\rho} \partial_\mu e^a_{\phantom{a}\nu} ,
\end{equation}
where $\bar{e}_a^{\phantom{a}\rho}$ denotes the inverse of
$e^a_{\phantom{a}\mu}$. Obviously, the connection
$\Gamma$ defined by these coefficients is not symmetric, i.e.
$\Gamma^\rho_{\mu\nu} \neq \Gamma^\rho_{\nu\mu}$.
Define then the torsion by formula
\begin{equation}
T^\rho_{\mu\nu} = \Gamma^\rho_{\mu\nu} - \Gamma^\rho_{\nu\mu} .
\end{equation}

In order to simplify the calculation, we consider the case of
spherical symmetry and choose the gauge fields
$e^a_{\phantom{a}\mu}$ as
\begin{equation}
e^a_{\phantom{a}\mu} = \mathrm{diag} \left( A, 1, 1, \frac{1}{A} \right)
,\quad
\bar{e}_a^{\phantom{a}\mu} = \mathrm{diag} \left( \frac{1}{A}, 1, 1, A \right) ,
\end{equation}
where $A = A(r)$ is a function depending only on the radial
coordinate $r$. Then, denoting the spherical coordinates on
$M$ by $(x^\mu) = (r, \vartheta, \varphi, t),\, \mu = 1, 2, 3, 0$,
the non-null components of the connection coefficients are
\begin{equation}
\Gamma^0_{10} = - \frac{A'}{A} ,\quad \Gamma^1_{11} = \frac{A'}{A} .
\end{equation}
It is easy to see that the only non-null components of the torsion are
\begin{equation}
T^0_{01} = - T^0_{10} = \frac{A'}{A} .
\end{equation}
Also, using the definitions \eqref{tildeR} and \eqref{R} of the curvatures, we obtain
\begin{equation}
\tilde{R}^0_{101} = - \tilde{R}^0_{110} = \frac{AA^{''} - 2A^{'2}}{A^2}
,\quad
R^\lambda_{\mu\nu\rho} = 0 \label{R_components}
\end{equation}
and all other components of $\tilde{R}^\lambda_{\mu\nu\rho}$ vanish.
In these expressions, we denote the first and second derivatives of
$A(r)$ by $A'$ and $A''$, respectively. The vanishing curvature
$R^\lambda_{\mu\nu\rho}$ agrees with the constraint \eqref{R_zero}.

Introduce then the noncommutativity parameters $\theta^{\mu\nu}$ and
suppose that we choose them to be
\begin{equation}
\left(\theta^{\mu\nu}\right) = \left(\begin{array}{cccc}
0 & 0 & 0 & \frac{1}{A(r)}\\
0 & 0 & b & 0\\
0 & -b & 0 & 0\\
-\frac{1}{A(r)} & 0 & 0 & 0\\
\end{array}\right) ,\label{theta}
\end{equation}
where $b$ is a nonvanishing constant.
Then we have
\begin{equation}
\tilde{\nabla}_1 \theta^{01} = - \tilde{\nabla}_1 \theta^{10} = 0
,\quad
\nabla_1 \theta^{01} = - \nabla_1 \theta^{10} = \frac{A'}{A^2} .
\end{equation}
This agrees with the constraint \eqref{tildenabla_symplectic} that
$\theta^{\mu\nu}$ is covariant constant under $\tilde{\nabla}$.

Finally, if we impose also the condition that the curvature
$\tilde{R}^{\mu\nu}_{\rho\sigma} = \theta^{\mu\lambda} \tilde{R}^\nu_{\lambda\rho\sigma}$ vanishes
(equivalent with \eqref{tildeR_zero}  due to \eqref{theta_invertible}), that implies $\nabla_\lambda \tilde{R}^{\mu\nu}_{\rho\sigma}$ vanishes too \eqref{tildeR_covariant_constant}, then from
\eqref{R_components} we obtain the following differential equation of
the second order for the unknown function $A(r)$:
\begin{equation}
AA^{''} - 2A^{'2} = 0 .\label{A_DE}
\end{equation}
The solutions of this equation is
\begin{equation}
A(r) = - \frac{1}{c_1 r + c_2} ,\label{A(r)}
\end{equation}
where $c_1$ and $c_2$ are two arbitrary constants of
integration. Therefore, in our simple example, the conditions necessary
to define a covariant star product on a symplectic manifold $M$
completely determine its connection. In addition, it is very
interesting to see that the covariant derivative of the torsion,
defined as
\begin{equation}
\nabla_\mu T^\nu_{\rho\sigma} = \partial_\mu T^\nu_{\rho\sigma} + \Gamma^\nu_{\mu\lambda} T^\lambda_{\rho\sigma} - \Gamma^\lambda_{\mu\rho} T^\nu_{\lambda\sigma} - \Gamma^\lambda_{\mu\sigma} T^\nu_{\rho\lambda} ,
\end{equation}
has the following non-null components
\begin{equation}
\nabla_1 T^0_{01} = - \nabla_1 T^0_{10} = \frac{AA^{''} - 2A^{'2}}{A^2} .
\end{equation}
Then, taking into account the equation \eqref{A_DE}, we conclude that the
torsion is covariant constant, $\nabla_\mu T^\nu_{\rho\sigma} =  0$,
a result which is in concordance with the condition \eqref{nabla_T_zero}.

Now we construct a noncommutative $U(2)$ gauge theory on the
space-time manifold $M$ presented in the previous example. Denote
the generators of the $U(2)$ group by $T_a, a=k,0$, with $k=1,2,3$;
here $T_k=\sigma_k$ (Pauli matrices) generates the $U(2)$ sector and
$T_0=I$ (the unit matrix) --- the $U(1)$ sector of the gauge group .
These generators satisfy the algebra \eqref{T_Lie_algebra}, where
only the structure constant $f^i_{jk}=2\epsilon_{ijk}$
($\epsilon_{ijk}$ are the totally antisymmetric Levi-Civita symbols)
of the $SU(2)$ sector are nonvanishing, the other components of
$f^a_{bc}$ being equal to zero. The anticommutator $\{T_a,
T_b\}=d^c_{ab}T_c$ also belongs to the algebra of $U(2)$, where
$d^0_{bc}=2\delta_{bc}, d^a_{b0}=2$ are the only nonvanishing
components.

We choose the 1-form gauge potential of the form \cite{Zet:2,Volkov}
\begin{equation}
A = u T_3 \rd t + w\left(T_2 \rd\vartheta - \sin\vartheta T_1 \rd\varphi\right) + \cos\vartheta T_3 \rd\varphi + v T_0 \rd t ,
\end{equation}
where $u,w,v$ are functions depending only on the radial coordinate $r$. We consider the metric $G_{\mu\nu}$ and its inverse $G^{\mu\nu}$ of the form
\begin{equation}
G_{\mu\nu} = \diag\left( \frac{1}{N}, r^2, r^2 \sin^2\vartheta, -N \right)
\end{equation}
and
\begin{equation}
G^{\mu\nu} = \diag\left( N, \frac{1}{r^2}, \frac{1}{r^2 \sin^2\vartheta}, -\frac{1}{N} \right) ,
\end{equation}
respectively, where $N$ is also a function depending only on $r$. For example, the following set of functions
\begin{equation}
u = u_0 + \frac{Q}{r} ,\quad w=0 ,\quad v=0 ,\quad N = 1-\frac{2M}{r}+\frac{Q^2+1}{r^2}
\label{uwvN}
\end{equation}
describes a colored black hole in the $SU(2)$ sector \cite{Volkov}. The metric $G_{\mu\nu}$ is of Reissner-Nordstr\"om type with electric charge $Q$ and unit magnetic charge \cite{Volkov}. It is the simplest solution of the Einstein-Yang-Mills field equations with a nontrivial gauge field.

We can obtain the noncommutative Yang-Mills field equations and their solutions by imposing the variational principle $\hat{\delta}\hat{S}_\mathrm{NCYM}=0$. However, it is simpler and equivalent to use the Seiberg-Witten map and determine the noncommutative gauge fields $\hat{A}_\mu$, the field strength $\hat{F}_{\mu\nu}$ and the metric $\hat{G}^{\mu\nu}$ order by order in the deformation parameter.

To this end, we denote the noncommutative quantities of our model by $\hat{\lambda}=\hat{\lambda}^a T_a$ (the gauge parameter), $\hat{A}=\hat{A}_\mu \dx{\mu}=\hat{A}_\mu^a T_a \dx{\mu}$ (the 1-form gauge potential) and $\hat{G}^{\mu\nu}=\hat{G}^{\mu\nu}I$ (the metric), and expand them as formal power series in $\theta$
\begin{align}
\hat{\lambda} &= \lambda + \lambda^{(1)} + \lambda^{(2)} + \cdots ,\nonumber\\
\hat{A}_\mu &= A_\mu + A^{(1)}_\mu + A^{(2)}_\mu + \cdots ,\\
\hat{G}^{\mu\nu} &= G^{\mu\nu} + G^{\mu\nu(1)} + G^{\mu\nu(2)} + \cdots ,\nonumber
\end{align}
where the zeroth order terms $\lambda$, $A_\mu$ and $G^{\mu\nu}$ are
the ordinary counterparts of $\hat{\lambda}$, $\hat{A}_\mu$ and
$\hat{G}^{\mu\nu}$, respectively. Using the Seiberg-Witten map for
the noncommutative gauge theory with the covariant star product
\cite{Chaichian:5} we obtain the following expressions for the first
order deformations:
\begin{align}
\lambda^{(1)} &= \frac{1}{4}\theta^{\rho\sigma}\left\{ \partial_\rho \lambda, A_\sigma \right\} ,\\
A_\mu^{(1)} &= -\frac{1}{4}\theta^{\rho\sigma}\left\{ A_\rho, \nabla_\sigma A_\mu + F_{\sigma\mu} \right\} ,\label{A_mu^(1)}\\
G^{\mu\nu(1)} &= -\theta^{\rho\sigma} A^0_\rho \nabla_\sigma G^{\mu\nu} .
\end{align}
Note that due to the particular form of the parameter
$\theta^{\mu\nu}$ \eqref{theta} and the solution \eqref{A(r)}, there
are in fact three noncommutativity parameters in our model: $c_1$,
$c_2$ and $b$. From now on we denote them by $c_1=\theta_1$ (of
dimension $T$), $c_2=\theta_2$ (of dimension $LT$) and $b=\theta_3$
(dimensionless).

The first order deformations of the field strength can be obtained from the definition \eqref{F_nc} by using \eqref{A_mu^(1)}:
\begin{equation}
F_{\mu\nu}^{(1)} = -\frac{1}{4}\theta^{\rho\sigma} \left( \left\{ A_\rho, \nabla_\sigma F_{\mu\nu} + D_\sigma F_{\mu\nu} \right\} - \left\{ F_{\mu\rho}, F_{\nu\sigma} \right\} \right) ,
\end{equation}
where
\begin{equation}
\nabla_\sigma F_{\mu\nu} = \partial_\sigma F_{\mu\nu} - \Gamma^\rho_{\sigma\mu} F_{\rho\nu} - \Gamma^\rho_{\sigma\nu} F_{\mu\rho}
\end{equation}
is the covariant derivative (it concerns the space-time manifold $M$) and
\begin{equation}
D_\sigma F_{\mu\nu} = \nabla_\sigma F_{\mu\nu} - i\left[ A_\sigma, F_{\mu\nu} \right]
\end{equation}
is the gauge covariant derivative (it concerns the gauge group).

In particular, for the colored black hole solution \eqref{uwvN} we obtain
\begin{align}
A_0^{a(1)} &= \left( 0, 0, \frac{1}{2}\sin(2\vartheta)\theta_3, -\frac{1}{2}\left(u_0+\frac{Q}{r}\right) \left[\left(u_0+\frac{3Q}{r}\right)\theta_1+\frac{2Q}{r^2}\theta_2\right] \right) ,\\
G^{\mu\nu(1)} &= 0 ,\\
F_{\mu\nu}^{0(1)} &= \left(\begin{array}{cccc}
0 & 0 & 0 & B(r)(r\theta_1+\theta_2) \\
0 & 0 & \cos(2\vartheta)\theta_3 & 0 \\
0 & -\cos(2\vartheta)\theta_3 & 0 & 0 \\
-B(r)(r\theta_1+\theta_2) & 0 & 0 & 0
\end{array}\right) ,
\end{align}
where $a=1,2,3,0$ and
\begin{equation}
B(r) = \frac{Q}{r^3}\left(2u_0 + \frac{3Q}{r}\right) ,
\end{equation}
and other components of $F_{\mu\nu}^{a(1)}$ are equal to zero. When the $U(1)$ sector is not empty $v(r)\neq 0$, we obtain a nonvanishing first order deformation of the metric $G^{\mu\nu(1)}\neq 0$, but for simplicity we prefer to consider only the colored black hole solution here.

All these first order deformations vanish if the noncommutativity parameters vanish $\theta_1, \theta_2, \theta_3 \rightarrow 0$. However, this limit cannot be achieved because the symplectic structure of the space-time $M$ imposes the condition $\det\left(\theta^{\mu\nu}\right)\neq 0$.

Finally, we mention that colored black holes and their generalizations with angular momentum and cosmological term, as well as solutions with cylindrical and plane symmetries have also been obtained \cite{Volkov}. It would be of interest to extend these results to the noncommutative theory.

\section{Conclusions and discussions}
\label{sec:conclusions}

We developed a noncommutative gauge theory by using a covariant star product
between differential forms on symplectic manifolds defined as in
\cite{McCurdy:1}. We followed the same way as in our recent paper \cite{Chaichian:2}, extending the
results of \cite{McCurdy:1} to the case of Lie-valued differential forms.

To simplify the calculations, we considered a space-time endowed only
with torsion. It has been shown that, in order to satisfy the
restrictions imposed by the associativity property of the covariant
star product, the torsion of the space-time has to be covariant
constant,  $\nabla_\mu T^\nu_{\rho\sigma} =  0$.
On the other hand, it has been argued that a covariant star product
defined in the case when the space-time is a symplectic manifold
endowed only with curvature is not possible. This is due to the
restrictions imposed by the associativity property of the covariant
star product which requires also the vanishing curvature. The
corresponding connection is therefore flat symplectic and this reduces
the applicability area of the covariant star product.

An illustrative example has been presented starting from the Poincar\'e
gauge theory. Using the gauge fields  $e^a_{\phantom{a}\mu}$
and fixing the gauge $\omega_\mu^{ab} = 0$ \cite{Zet} we defined the
nonsymmetric connection
$\Gamma^\rho_{\mu\nu} = \bar{e}_a^{\phantom{a}\rho} \partial_\mu e^a_{\phantom{a}\nu}$.
We deduced that, in this case, the conditions necessary to define
a covariant star product on a symplectic manifold $M$ completely
determine its connection.

Some other possible applications of this covariant star product have
been also analyzed. First, it will be very important to generalize the
Seiberg-Witten map to the case when the ordinary derivatives are
replaced with covariant derivatives and the Moyal star product is replaced by the
covariant one. Second, we can try to develop a noncommutative gauge
theory of gravity considering the symplectic manifold $M$ as the
background space-time. For such a purpose, we have to verify if the
noncommutative field equations do not impose too many restrictive
conditions on the connection $\Gamma^\rho_{\mu\nu}$,
in addition to those required by the existence of the covariant
product. However, the problem of which gauge group we can choose
remains unsolved. The Poincar\'e group can not be used because it
does not close with respect to the star product. A possibility will be to
choose the group $GL(2, \mathbb{C})$, but in this case
we obtain a complex theory of gravitation \cite{Stern,Chamseddine}. Another
possibility is to consider the universal enveloping of the Poincar\'e
group, but this is infinite dimensional and we must find criteria to
reduce the number of the degrees of freedom to a finite one. Some possible
ideas are given for the case of $SU(N)$ or GUT theories
in \cite{Calmet}, where it is argued that the infinite number of
parameters can in fact all be expressed in terms of right number of
classical parameters and fields via the Seiberg-Witten maps.

Two ways to generalize the Poisson bracket and the covariant star product to the algebra of tensor fields on a symplectic manifold have been proposed recently \cite{Chaichian:6,Vassilevich:2}. In the latter work one studied covariant star products on spaces of tensor fields defined over a Fedosov manifold with a given symplectic structure and a given flat torsionless symplectic connection. In the former approach one considers such generalizations on symplectic manifolds and Poisson manifolds that impose as few constraints as possible on the connections. Both of these approaches are in part motivated by their possible application to noncommutative gravity.

\paragraph{Acknowledgements.} The support of the Academy of Finland under
the Projects No. 121720 and No. 127626 is gratefully acknowledged. The
work of M. O. was fully supported by the Jenny and Antti Wihuri
Foundation. G. Z. acknowledges the support of CNCSIS-UEFISCSU Grant
ID-620 of the Ministry of Education and Research of Romania.

\appendix
\section{Appendix}
\label{sec:appendix}

 We verify the associativity property up to the second order in $\hbar$.
For simplicity we denote the exterior product by
$\alpha \wedge \beta = \alpha \beta$.

Introducing \eqref{star_product} into \eqref{associativity} we obtain successively
\begin{multline}
\left[ \alpha \beta + \frac{i\hbar}{2} C_1 (\alpha, \beta) + \left(\frac{i\hbar}{2}\right)^2 C_2 (\alpha, \beta) + \cdots \right] \star \gamma \\
= \alpha \star \left[ \beta \gamma + \frac{i\hbar}{2} C_1 (\beta, \gamma) + \left(\frac{i\hbar}{2}\right)^2 C_2 (\beta, \gamma) + \cdots \right]
\end{multline}
or
\begin{align*}
(\alpha \beta) \gamma &+ \frac{i\hbar}{2} \left[ C_1 (\alpha \beta, \gamma) + C_1 (\alpha, \beta) \gamma \right] \\
&+ \left(\frac{i\hbar}{2}\right)^2 \left[ C_2 (\alpha \beta, \gamma) + C_1 (C_1 (\alpha, \beta), \gamma) + C_2 (\alpha, \beta) \gamma \right] + \cdots \\
&= \alpha (\beta \gamma) + \frac{i\hbar}{2} \left[ C_1 (\alpha, \beta \gamma) + \alpha C_1 (\beta, \gamma) \right] \\
&+ \left(\frac{i\hbar}{2}\right)^2 \left[ C_2 (\alpha, \beta \gamma) + C_1 (\alpha, C_1 (\beta, \gamma)) + \alpha C_2 (\beta, \gamma) \right] + \cdots .
\end{align*}
Identifying the terms of different orders in $\hbar$ we obtain
\begin{gather}
(\alpha \beta) \gamma = \alpha (\beta \gamma) ,\label{associativity_zero_order}\\
C_1 (\alpha \beta, \gamma) + C_1 (\alpha, \beta) \gamma = C_1 (\alpha, \beta \gamma) + \alpha C_1 (\beta, \gamma) ,\label{associativity_1st_order}\\
C_2 (\alpha \beta, \gamma) + C_1 (C_1 (\alpha, \beta), \gamma) + C_2 (\alpha, \beta) \gamma = C_2 (\alpha, \beta \gamma) + C_1 (\alpha, C_1 (\beta, \gamma)) + \alpha C_2 (\beta, \gamma) .\label{associativity_2nd_order}
\end{gather}
\eqref{associativity_zero_order} is verified because the exterior product has this property.

Using \eqref{C_1}, the \eqref{associativity_1st_order} becomes
\[
\theta^{\mu\nu} \left[ \nabla_\mu (\alpha \beta) (\nabla_\nu \gamma) + (\nabla_\mu \alpha) (\nabla_\nu \beta) \gamma -  (\nabla_\mu \alpha) \nabla_\nu (\beta \gamma) - \alpha (\nabla_\mu \beta) (\nabla_\nu \gamma) \right] = 0
\]
that is satisfied due to the Leibniz rule $\nabla_\mu (\alpha \beta) = (\nabla_\mu \alpha) \beta + \alpha (\nabla_\mu \beta)$.

In order to verify \eqref{associativity_2nd_order} we write it as
\begin{equation}
\begin{split}
\delta C_2 (\alpha, \beta, \gamma) &\equiv C_2 (\alpha, \beta \gamma) - C_2 (\alpha, \beta) \gamma - C_2 (\alpha \beta, \gamma) + \alpha C_2 (\beta, \gamma) \\
&= \{ \{ \alpha, \beta \}, \gamma \} - \{ \alpha, \{ \beta, \gamma \} \} \label{delta_C_2}
\end{split}
\end{equation}
We calculate the right-hand side of \eqref{delta_C_2} first
\begin{multline}
\{ \{ \alpha, \beta \}, \gamma \} - \{ \alpha, \{ \beta, \gamma \} \} = (-1)^{|\alpha| (|\beta| + |\gamma|)} \{ \beta, \{ \gamma, \alpha \} \} \\
= (-1)^{|\alpha| (|\beta| + |\gamma|)} \theta^{\mu\nu} (\nabla_\mu \beta) \nabla_\nu \left( \theta^{\rho\sigma} (\nabla_\rho \gamma) (\nabla_\sigma \alpha) \right)
= - \theta^{\mu\nu} (\nabla_\nu \theta^{\rho\sigma}) (\nabla_\rho \alpha) (\nabla_\mu \beta) (\nabla_\sigma \gamma) \\
+ \theta^{\mu\nu} \theta^{\rho\sigma} \left[ (\nabla_\mu \nabla_\rho \alpha) (\nabla_\nu \beta) (\nabla_\sigma \gamma) - (\nabla_\mu \alpha) (\nabla_\rho \beta) (\nabla_\sigma \nabla_\nu \gamma) \right] ,\label{double_Pbs}
\end{multline}
where the graded symmetry property \eqref{Moyal_symmetry} (for $n = 1$) and the graded Jacobi identity \eqref{Jacobi_Pb} of the Poisson bracket are used in the first equality,
the expression \eqref{C_1} for the Poisson bracket in the second step, and the symmetry properties of $\theta^{\mu\nu}$ and the exterior product,
$\alpha \beta = (-1)^{|\alpha| |\beta|} \beta \alpha$, in the last equality.
Then we introduce the decomposition of the second covariant derivative
of an arbitrary differential form as
\begin{equation}
\nabla_\mu \nabla_\rho \alpha = \frac{1}{2} \left\{ \nabla_\mu, \nabla_\rho \right\} \alpha - \frac{1}{2} T^\lambda_{\mu\rho} \nabla_\lambda ,
\end{equation}
implied by \eqref{cocd} and \eqref{R_zero}, into \eqref{double_Pbs}. Finally, using the cyclic property \eqref{T_cyclic} for the torsion, we obtain
\begin{multline}
\{ \{ \alpha, \beta \}, \gamma \} - \{ \alpha, \{ \beta, \gamma \} \}
= - \left( \theta^{\mu\sigma} \nabla_\sigma \theta^{\nu\rho} + \frac{1}{2} \theta^{\nu\sigma} \theta^{\rho\lambda} T^\mu_{\sigma\lambda} \right) (\nabla_\nu \alpha) (\nabla_\mu \beta) (\nabla_\rho \gamma) \\
+ \frac{1}{2} \theta^{\mu\nu} \theta^{\rho\sigma} \left[ \left\{ \nabla_\mu, \nabla_\rho \right\} \alpha (\nabla_\nu \beta) (\nabla_\sigma \gamma) - (\nabla_\mu \alpha) (\nabla_\rho \beta) \left\{ \nabla_\nu , \nabla_\sigma\right\} \gamma \right] ,\label{double_Pbs_final}
\end{multline}

Next, we calculate the left-hand side of \eqref{delta_C_2}, the Hochschild coboundary of $C_2$ (see \cite{McCurdy:1} for details). First we calculate $C_2 (\alpha, \beta \gamma) - C_2 (\alpha, \beta) \gamma$ and $C_2 (\alpha \beta, \gamma) - \alpha C_2 (\beta, \gamma)$, then substracting them yields
\begin{multline}
\delta C_2 (\alpha, \beta, \gamma) = \frac{1}{2} \theta^{\mu\nu} \theta^{\rho\sigma} \left[ (\nabla_\mu \nabla_\rho \alpha) 2(\nabla_{(\nu} \beta \nabla_{\sigma)} \gamma) - 2(\nabla_{(\mu} \alpha \nabla_{\rho)} \beta) (\nabla_\nu \nabla_\sigma \gamma) \right] \\
+ \frac{1}{3} \left( \theta^{\mu\sigma} \nabla_\sigma \theta^{\nu\rho} + \frac{1}{2} \theta^{\nu\sigma} \theta^{\rho\lambda} T^\mu_{\sigma\lambda} \right) \left( (\nabla_\rho \alpha) 2(\nabla_{(\mu} \beta \nabla_{\nu)} \gamma) - 2(\nabla_{(\mu} \alpha \nabla_{\nu)} \beta) (\nabla_\rho \gamma) \right) ,
\end{multline}
where we denote $\nabla_{(\mu} \alpha \nabla_{\rho)} \beta = \frac{1}{2} \left[ (\nabla_\mu \alpha) (\nabla_\rho \beta) + (\nabla_\rho \alpha) (\nabla_\mu \beta) \right]$. By using the symmetries of the factor $\theta^{\mu\nu} \theta^{\rho\sigma}$ and the cyclic relation implied by \eqref{Jacobi_covariant} and \eqref{T_cyclic},
\begin{equation}
\sum_{(\mu, \nu, \rho)} \left( \theta^{\mu\sigma} \nabla_\sigma \theta^{\nu\rho} + \frac{1}{2} \theta^{\nu\sigma} \theta^{\rho\lambda} T^\mu_{\sigma\lambda} \right) = 0 ,
\end{equation}
we find
\begin{multline}
\delta C_2 (\alpha, \beta, \gamma) = \frac{1}{2} \theta^{\mu\nu} \theta^{\rho\sigma} \left[ \left\{ \nabla_\mu, \nabla_\rho \right\} \alpha (\nabla_\nu \beta) (\nabla_\sigma \gamma) - (\nabla_\mu \alpha) (\nabla_\rho \beta) \left\{ \nabla_\nu, \nabla_\sigma \right\} \gamma \right] \\
- \left( \theta^{\mu\sigma} \nabla_\sigma \theta^{\nu\rho} + \frac{1}{2} \theta^{\nu\sigma} \theta^{\rho\lambda} T^\mu_{\sigma\lambda} \right) (\nabla_\nu \alpha) (\nabla_\mu \beta) (\nabla_\rho \gamma) ,
\end{multline}
This is the same result as in \eqref{double_Pbs_final} and therefore
we have verified the associativity of our star product to the second
order in $\hbar$.

\end{document}